# A model for system developers to measure the privacy risk of data

Awanthika Senarath
School of Engineering and IT
University of New South Wales
Australia
a.senarath@student.unsw.edu.au

Marthie Grobler
CSIRO, Data61
Australia
marthie.grobler@data61.csiro.au

Nalin A. G. Arachchilage
School of Engineering and IT
University of New South Wales
Australia
nalin.asanka@adfa.edu.au

## Abstract

*In this paper, we propose a model that could be used by system developers to measure the privacy risk perceived by users when they disclose data into software systems. We first derive a model to measure the perceived privacy risk based on existing knowledge and then we test our model through a survey with 151 participants. Our findings revealed that users' perceived privacy risk monotonically increases with data sensitivity and visibility, and monotonically decreases with data relevance to the application. Furthermore, how visible data is in an application by default when the user discloses data had the highest impact on the perceived privacy risk. This model would enable developers to measure the users' perceived privacy risk associated with data items, which would help them to understand how to treat different data within a system design.*

## 1. Introduction

The new European General Data Protection Regulations (GDPR) that came into effect in 2018 has generated considerable interest towards privacy design guidelines in software system designers, such as the Privacy by Design (PbD) principles [1]. However, PbD has been criticized for its limitations in being incompatible and different to the usual activities developers perform when they design software systems [2, 3]. One such limitation that has been frequently raised is its lack of support for developers to understand users' perceived privacy risk when they design software systems [3]. Lack of understanding on the privacy risk perceived by users could result in systems that do not cater for user privacy expectations and hence invade user privacy when users interact with those systems [4]. That is, users perceive a privacy risk when they disclose data into software systems, depending on the system they interact with and the data they are required to disclose [5]. However, heretofore developers are oblivion to this privacy risk users perceive [4, 6, 7]. If a system collects and stores data users perceive to have a higher privacy risk without anonymity or encryption, these data could be hacked and used by other parties, which could result in cyber bullying and identity theft. Because of this, software systems continue to fail user privacy even with the strict privacy regulations [1] and numerous privacy guidelines such as PbD that guide developers to embed privacy into software systems [4].

In this research we propose a model that would enable software developers to understand the privacy risk users perceive when they disclose data into software systems. Previous research has shown that the knowledge of the properties of data (such as how sensitive the content is and how visible the content is in a system) could be used [8] to measure the privacy risk of content in software systems. Consequently, it has been identified that this perceived privacy risk can be identified through the data disclosure decisions made by users [5]. For example, how sensitive data is, and how relevant data is to the application, are known to have an effect on the data disclosure decisions made by users [5]. This is because users' data disclosure decisions are closely related to their perceived privacy risk [9–11].

Building on this knowledge, we first measure users' perceived privacy risk through their data disclosure decisions, and model the perceived privacy risk using the properties of the data such as data sensitivity and visibility. Then, using a survey with 151 respondents we observe how good our model fits with the actual privacy risk perceived by users. Our findings disclosed that visibility of data has a significant impact on the privacy risk perceived by users. We also observed that the relatedness of data to the purpose of the application, has a negative impact on the privacy risk perceived by users when they interact with systems. With these findings developers can understand how they need to design systems to reduce the risk of data items within systems. This would eventually lead to systems that respect user privacy.

The paper is structured as follows. We first discuss the background of perceived privacy risk, and privacy risk measurement to establish the grounds on which our work stand. Then, building on the existing theoretical knowledge on measuring privacy risk, we first logically build our model to measure users' perceived privacy risk associated with disclosing data items in a given software system setting. Thereafter, we describe the experiment we conducted to measure the actual privacy risk perceived by users when they disclose their data. Next, we present our results where we show how good out model fits the observations, followed by a discussion of the observed variations and limitations of our model. Finally, we present our conclusions.

## 2. Background

Our focus in this research is to develop a metric to measure the users' perceived privacy risk associated with data (such as their name, address and email address) in software systems (such as their banking app, their social networking account etc.). It has been identified that understanding the data disclosure decisions made by users when interacting with software systems could help understanding their perceived privacy risk [9–11]. Nevertheless, among many research studies that attempts to interpret users privacy risk and their data disclosure decisions [5, 10, 12–14], so far no attempt has been made to measure this privacy risk perceived by users when they disclose data into systems, in a comprehensive way to software developers.

Most research that observe disclosure decisions of users attempt to identify factors that could increase data disclosure. For example, focusing on the intrinsic properties of the data being shared, Bansal et al. have shown that users' intention to disclose health information is affected by the sensitivity of the data [15]. This intrigued our interest. Malhotra et al. have also shown that consumer willingness to share personal data in commercial platforms is affected by the sensitivity of the data [11]. Similarly, Malheiros et al. [5] have shown that sensitivity of data items such as date of birth and occupation had a significant affect on the decisions of the users to disclose that data into software systems. However, how these parameters correlate when users make their decisions to disclose data and how software developers could make use of this information when they design software systems are not yet known.

Consequently, it is said that users are more likely to disclose data when they are shown the decisions made by their friends [14] or other users [16]. Similarly, Acquisti et al. found that changing the order of intrusiveness of the data being requested also makes users disclose more data when interacting with software systems [17]. Furthermore, testing the effect of the justification provided by the system when requesting data Knijnenburg and Kobsa [18] revealed that when users are told *this data is useful for you* users are more likely to disclose data to the application. Nevertheless, these research focus either on the features of the system that requests data [5, 10, 13] or the personality of the user who discloses data [19] and attempt to find ways to increase user data disclosure [14]. We approach this problem from a different standpoint. We believe that rather than finding ways to increase data disclosure, developers should implement better privacy in systems and transparently communicate with users so that the cumulative privacy risk in systems would be reduced. For this developers should be able to measure and understand the privacy risk perceived by users when they disclose data.

From a perspective of privacy risk measurement, Maximilien et al. [8] have shown that a metric for privacy risk in a given context can be obtained by multiplying the measurement for sensitivity of a data item with the measurement for visibility the data item gets in an application. They define their metric for privacy risk as "a measurement that determines their [the user's] willingness to disclose information associated with this item" [8]. Using this metric, Minkus and Memon [20] have attempted to measure the privateness of Facebook users from their privacy settings. However, privacy risk is a contextual measurement. The context in which data is being disclosed [19, 21] is known to have an effect on user disclosure decisions [5, 12]. For example, it is said that users have a negative attitude towards rewards for data disclosure when the requested data appears irrelevant for a system [10], whereas they accepted the rewards if the data is relevant for the system. However, the current model by Maximilien et al. [8] for privacy risk measurement of content, does not account for the relatedness of data. Nevertheless, when a developer attempts to make use of the perceived privacy risk of data to support him in the decisions to embed privacy into the system (for example embedding data minimization into a software system), how relevant the data is to the system is important [6]. The requirements established by the recent reforms in the GDPR to collect only relevant data, and communicate the use of data to system users [1] exacerbates the importance of developers accounting for data relatedness when designing privacy-respectful software systems.

In this research, we focus on the effect of data sensitivity, the relevance of the data for an application and the visibility the data gets in the application on the privacy risk perceived by users. With this we propose a model that could communicate the effect of data sensitivity, visibility and the relatedness of data for a particular application on the privacy risk perceived by users to software developers and privacy researchers. By software developers, we refer to all those who are involved in making the decisions on collecting data, designing and implementing software systems. The proposed model would help them to understand and incorporate privacy risk perceived by users into the software system designs and assist the development of privacy respectful software systems.

## 3. Study Design

In this section we first introduce the parameters of data we are interested in. Then using these parameters we derive and propose a model to measure privacy risk of data items based on existing theoretical knowledge.

The goal of our research was to develop a measurement to calculate the privacy risk perceived by users when they disclose data into software systems. Referring to previous research we identified data sensitivity (S), relatedness (R) and visibility (V) of data has an impact on the privacy risk that is perceived by users when they make data disclosure decisions. For the context of this research we define data sensitivity, visibility and relatedness of data to be parameters that depend only on the data item $D_i$ and the application context in which it is being accessed/used $C_j$. Following subsections define the parameters for the context of the model we propose.

### 3.1. Data Sensitivity

We define the sensitivity of a particular data item to be a parameter that is dependent on the data item $D_i$ itself. That is credit card number is inherently more sensitive for a user than their age. We define sensitivity of a data item to be the perceived impact of loss of that particular data item. We define sensitivity in three categorical values based on logical reasoning and the definition of sensitive data in the European Data Protection Regulations (GDPR) [1]. Three categories are considered to be cognitively more manageable than complex scales with more levels of categorization [7]. Our definitions for categorization is given in table 1.

### 3.2. Data Visibility (V)

We define the visibility of a data element to be an inherent property gained by a particular data element $D_i$ in a particular application context $C_j$ due to the design of the application. That is how visible the data item would be by default once the user disclose the data item to the application. If the application by default allows the data to be seen only by the user, we define that data item has the lowest visibility. These categories are defined on the basis of the survey conducted by Minkus and

**Table 1.** Data Sensitivity (S)

| Category | Description | S |
|---|---|---|
| Category I - Highest sensitivity | Data that could be used to identify a unique characteristic of a person. For example, a person's race, religion or HIV status. | 3 |
| Category II - Moderate sensitivity | Personally Identifiable information about the person. For example, a person's name, address, mobile number. | 2 |
| Category III - Low sensitivity | Any other detail about a person that may have an impact of loss, however, would not affect the person. For example, a person's high school. | 1 |

Memon [20]. In their attempts to scale Facebook privacy settings according to their visibility, they have asked participants questions that investigate the users' perception of visibility of their content in Facebook. Building on their reasoning we logically form the three visibility categories presented in Table 2.

**Table 2.** Data Visibility

| Category | Description | V |
|---|---|---|
| Category I - Highest visibility | Data would be seen by any one by default. Data is visible in the application by default. For example the name of a user in Facebook. | 3 |
| Category II - Moderate visibility | Data would be seen by a controlled set of users by default. For example, content that can be only see by the friends of the user in Facebook. | 2 |
| Category III - Low visibility | Data would be seen by any one by default. Data is visible in the application by default. For example, your pin number in the banking app will not be visible to any one. | 1 |

### 3.3. Data Relatedness (R)

We define the relatedness of a data element $D_i$ to be a property that is defined by the application context $C_j$. That is based on the requirements of the application, the data could be highly related to the application (for example, your bank account number for your banking application) or not related at all. This is determined by the primary functionality of the application defined by the application requirements. We build this categorization based on logical reasoning. While it has been widely accepted that the relatedness of data affects the privacy risk perceived by users when they disclose data into software systems [19, 21], so far there is no evi-dence as to how related a data item should be in order to make users feel comfortable sharing those data into the system. Therefore, based on logical reasoning, we propose the categorization present in table 3 for scaling data relatedness to a software system.

**Table 3. Data Relatedness**

| Category | Description | R |
|---|---|---|
| Category I - Highest relatedness | Data the application cannot do without. These data are absolutely necessary for the primary functionality of the application. | 3 |
| Category II - Moderate relatedness | Data could add additional functionality to the application. For example, data that could deliver benefits through data analysis techniques. | 2 |
| Category III - Low relatedness | Data the application can do without. For example, data that is not needed for the functionality of the application. | 1 |

According to our definitions presented in tables 1-3, the relatedness of a data element $D_i$ in an application context $C_j$ also takes categorical values $R_{i,j} \in \{1,2,3\}$, visibility of a data element $D_i$ in an application context $C_j$ takes categorical values $V_{i,j} \in \{1,2,3\}$ and the sensitivity of a data element $D_i$ takes categorical values $S_i \in \{1,2,3\}$.

### 3.4. A Model to calculate privacy risk of data elements :

In order to model users' perceived privacy risk, we define the calculated privacy risk $P_{i,j}$ of a data element $D_i$ in an application context $C_j$ as follows. Building up on the relationship proposed by Maximilien et al. [8] we define that the privacy risk $P_{i,j}$ of a data element $D_i$ in an application context $C_j$ monotonically increases with the sensitivity of a data item $S_i$ and the visibility of a data item in a given context $V_{(i,j)}$. This has been previously used by Minkus and Memon [20] in determining the privacy level of Facebook privacy settings for a particular user. Then, we propose that the privacy risk $P_c$ of a data element $D_i$ in an application context $C_j$ is in a monotonically decremental relationship with the relatedness of the data element $D_i$ to the application context $C_j$. This is based on the knowledge that users perceive low privacy risk when disclosing data items that are relevant to the application as opposed to data elements that do not appear relevant [18]. Therefore, we propose that an approximation for the privacy risk $P_{i,j}$ of a data element $D_i$ in an application context $C_j$ can be obtained by,

$$\text{Privacy Risk } P_{(i,j)} = \frac{S_i^a \times V_{(i,j)}^b}{R_{(i,j)}^c}$$

where a,b and c values could take any real number. However, as we are aiming for an approximation we limit a,b,c to whole numbers.

According to this calculation Privacy Risk $P_{(i,j)}$ of a data element $D_i$ in an application context $C_j \in \{x/x \text{ in IR where, } 0 < x\}$. Next, in order to see how closely the proposed model fit the actual privacy risk perceived by the users when they disclose data we conducted a survey study.

## 4. Research Study

Our goal in conducting the research study is to observe how close the relationship we proposed using data sensitivity, visibility and relatedness approximate the actual privacy risk perceived by users. Building on the work of Maximilien et al. [8] we define perceived privacy risk $P_{i,j}$ to be "a measurement that determines the user's feeling of discomfort in disclosing an data item $D_i$ in an application context $C_j$". We conducted two separate user studies for this research.

### 4.1. Study I :

The first study was an online survey with 151 internet users to obtain the dependent variable of our model, the privacy risk perceived by users $P_{i,j}$. Users' perceived privacy risk can be interpreted as their discomfort or reluctance for data disclosure in software systems [9–11]. Therefore, in the user survey we obtained the discomfort of users when they disclose data into software systems $F_d$ as a measurement of their perceived privacy risk $P_{i,j}$. For this we defined three data disclosure scenarios.

- Health-care application that allows remote consultancy with doctors - with data being visible to the user and the doctor.

- Social Networking application - with no control over data visibility (Cannot control who can view the data once disclosed).

- Banking application - with the data being visible only to the user (and the bank).

We defined these scenarios with three different visibility levels based on our definitions in table 2. We used ten data items including demographic data and sensitive data following the GDPR [1]. The data items we provided are name, age, address, mobile number, email address, occupation, blood type, credit card number, medicine taken, and birthday. We asked the participants how they would feel if they are to disclose these 10 data items in the four application contexts. We define a five point Likert scale to express their *feeling of disclosure* $F_d$, with values, very uncomfortable, somewhat uncomfortable, neutral, somewhat comfortable and very comfortable. We alternatively used reverse ordered Likert scales to ensure the validity of the answers. We consider $F_d$ to be a function of the sensitivity of the data item i ($S_i$), visibility of the data item in the application j ($V_{i,j}$) and the relatedness of the data item to the context of the software system j ($R_{i,j}$). Our goal is to determine how close the calculated privacy risk from the model we proposed $P_{i,j}$ would approximate $F_d$.

In the survey we also included an open ended question in the questionnaire to further observe the reasons for the difference in the feeling of discomfort ($F_d$) users expressed. With this we aimed to obtain further insights as to why users demonstrate different discomfort levels when they disclose different data items into different application contexts. At the end of the survey, we included questions to extract the demographics of the participants which is presented in table 4.

Table 4. Participants (151)

| Gender | No. of Participants |
|---|---|
| Male | 87 |
| Female | 64 |
| **Education** | |
| Completed School Education | 5 |
| Professional Diploma | 9 |
| Bachelor's Degree | 87 |
| Masters/PhD | 50 |
| **Age** | |
| 18-24 | 31 |
| 25-32 | 101 |
| 33-40 | 13 |
| 41 or above | 6 |

The survey design was evaluated with two participants (graduate students in the university not connected to the research). We fine tuned the wording of the questionnaire with the feedback of these two participants. Then the survey was distributed using social media platforms (Facebook, LinkedIn and Twitter) and personal connections of the authors. The research methodology (survey design, participant recruitment and results collection) was approved by the university ethic committee responsible for ethical conduct of studies that involve human subjects.

In the invitation email we sent to participants, we included a brief introduction about the survey and the duration of the survey (under 10 minutes, calculated using the participants who evaluated the questionnaire). We provided the participants with the contact details of the researchers in case they wanted to contact us for more information. Before proceeding with the survey participants were given an introduction to the survey with details about the survey and the type of data we collect. We also informed the participants that they could exit the survey at any time without submitting their answers. Participants were asked to proceed with the survey if they give us (the researchers) consent to collect and store the details they submit with the survey.

We measured the participant adequacy while collecting data and stopped data collection when we reached sample adequacy at KMO = 0.8 (A KMO value 0.8 is considered good in calculating correlations [22]). We had 157 responses at that point. We then analyzed the data and eliminated 6 responses that were either incomplete or invalid as the participant had selected the same choice in the Likert scale for all options.

To transform the likert scale input into a measurement of the feeling of discomfort of the participants, we assigned values from 1 to 5 for the answers we received on the Likert scale as follows. Very Comfortable (1), Somewhat Comfortable (2), Neutral (3), Somewhat Uncomfortable (4), Very Uncomfortable (5). Through this we obtained $F_d \in \{1,2,3,4,5\}$ of users for the 30 scenarios (ten data items in three application contexts) that represent the user's feeling of discomfort in disclosing data.

### 4.2. Study II :

The second study was a focus group with 4 software developers, to obtain the independent variables of the model (sensitivity, visibility and relatedness) for the three data disclosure scenarios we used in the survey. As our goal is to introduce a metric for software developers to evaluate the privacy risk perceived by users, we calculated $P_{(i,j)}$ through a focus group with 4 participants with software development experience. We believe this approach would closely represent the context in which software developers would discuss and evaluate the sensitivity, visibility and the relatedness of the data elements they use in software systems, at design stage. The focus group took 40 minutes and the participants were volunteers.

In the focus group we first discussed the data items as individual elements and categorize them according to sensitivity. For this we provided the participants with the three categorical definitions we defined in table 1. Next, for all three application scenarios, we asked the developers to categorize the ten data items according to their relatedness to the application context and provided them with table 3. We encouraged the participants to raise arguments and discuss and clarify different opinions in categorizing data. As visibility was predetermined when we defined the three application scenarios in the survey and communicated to users in the user study we did not evaluate it here. During the focus group, we also evaluated our model for data categorization presented in Table 1, 2 and 3. We encouraged the participants to argue and raise any concerns they had on the three categories we defined and their appropriateness in categorizing the data. We discuss the concerns raised by the participants in the focus group when we discuss our findings.

**4.2.1. Data Analysis** After obtaining the S,V,R combinations for the 30 scenarios from the focus group, we tested the calculated perceived privacy risk using our model against the perceived privacy risk values we obtained through the user study. We first attempted to fit our model on the raw data available (151 users and 30 instances, altogether 4530 instances). However, due to the relatively high variation of data, it was not possible to fit a model to the data set. That is, the same combination of S,V, R values had multiple perceived privacy risks varying from 1 to 5. This is expected because users have very different perceived privacy risks. We then averaged the perceived privacy risk of all 151 users to obtain 30 distinct mean perceived privacy risk values for the 30 scenarios tested. Then we used these values to observe the goodness of fit of our proposed model in Matlab.

We used qualitative methods to analyse the answers to the open ended question using two independent coders. We followed the grounded theory approach where the coders coded data by eliciting codes from the data available without any prejudice [23]. This was done in NVivo [24]. Coders reached code saturation at 49 and 103 respectively. The two coders came up with 6 common codes and 7 and 20 codes present in either of the coders at the end. This was because one coder had very granular level codes while the other coder had coded data at a higher level. For example, one coder had a code saying *visibility of data*, while the other coder had three separate codes for the same content as *controlling who can see my data*, *application providing tools to hide data from public* and *controlling data in the app*. Then both coders iteratively evaluated their codes and merged similar codes together to come up with 11 final codes that explain the differences in the privacy risks perceived by the participants.

## 5. Results

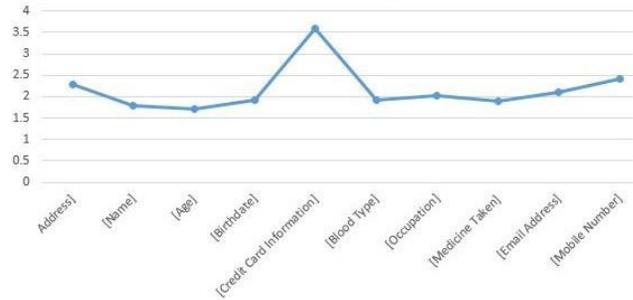

**Figure 1.** Feeling of Discomfort in Disclosure - Health application

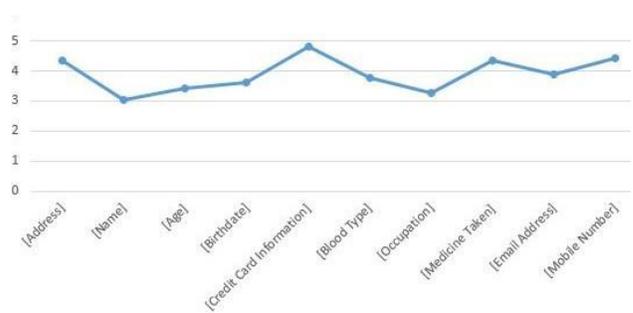

**Figure 2.** Feeling of Discomfort in Disclosure - Social Networking application

We tested the validity of our results with Cronbach's alpha (0.91) (a Cronbach's alpha > 0.7 is considered acceptable [25]) and the participant adequacy for correlations with KMO (KMO = 0.8269). The charts (figures 1-3) shows the averages of the disclosure feeling of the 151 participants on the 10 data items across the three scenarios. It can be seen that in all scenarios except for the banking app users had the highest discomfort in sharing their credit card information, and this was followed by medical information except for the medical application, which suggests users feel higher risk when disclosing sensitive data yet, it was reduced when they felt that the data was related to the application.

**Table 5.** Model Fitting - basic model

| Model | a (95% CI) | Goodness of fit | | |
|---|---|---|---|---|
| | | SSE | $R^2$ | RMSE |
| $\frac{S_i^1 \times V_{(i,j)}^1}{R_{(i,j)}^1}$ | 0.24 | 67.8 | 0.6 | 1.5 |

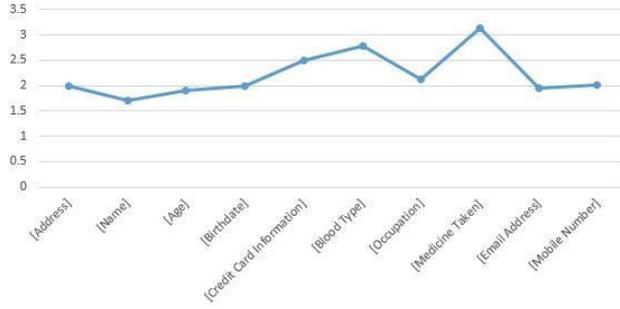

**Figure 3. Feeling of Discomfort in Disclosure - Banking application**

**Table 6. Model Fitting**

| Model | a (95% CI) | Goodness of fit | | |
|---|---|---|---|---|
| | | SSE | $R^2$ | RMSE |
| $a(\frac{S_i^1 \times V_{(i,j)}^1}{R_{(i,j)}^2})$ | 0.24 | 15.22 | 0.4353 | 0.7373 |
| $a(\frac{S_i^1 \times V_{(i,j)}^1}{R_{(i,j)}^3})$ | 0.21 | 16.7 | 0.3803 | 0.7723 |
| $a(\frac{S_i^1 \times V_{(i,j)}^2}{R_{(i,j)}^1})$ | 0.10 | 9.335 | 0.6536 | 0.5774 |
| $a(\frac{S_i^1 \times V_{(i,j)}^2}{R_{(i,j)}^2})$ | 0.08 | 12.57 | 0.5336 | 0.67 |
| $a(\frac{S_i^1 \times V_{(i,j)}^2}{R_{(i,j)}^3})$ | 0.08 | 14.4 | 0.4657 | 0.7171 |
| $a(\frac{S_i^1 \times V_{(i,j)}^3}{R_{(i,j)}^1})$ | 0.03 | 8.285 | 0.6926 | 0.544 |
| $a(\frac{S_i^1 \times V_{(i,j)}^3}{R_{(i,j)}^2})$ | 0.03 | 11.73 | 0.5646 | 0.5646 |
| $a(\frac{S_i^1 \times V_{(i,j)}^3}{R_{(i,j)}^3})$ | 0.02 | 13.71 | 0.4912 | 0.6998 |
| $a(\frac{S_i^2 \times V_{(i,j)}^1}{R_{(i,j)}^1})$ | 0.08 | 13.94 | 0.4828 | 0.7055 |
| $a(\frac{S_i^2 \times V_{(i,j)}^1}{R_{(i,j)}^2})$ | 0.07 | 15.38 | 0.4294 | 0.7411 |
| $a(\frac{S_i^2 \times V_{(i,j)}^1}{R_{(i,j)}^3})$ | 0.07 | 16.45 | 0.3895 | 0.7666 |
| $a(\frac{S_i^2 \times V_{(i,j)}^2}{R_{(i,j)}^1})$ | 0.03 | 11.06 | 0.5897 | 0.6284 |
| $a(\frac{S_i^2 \times V_{(i,j)}^2}{R_{(i,j)}^3})$ | 0.02 | 14.78 | 0.4515 | 0.7266 |
| $a(\frac{S_i^2 \times V_{(i,j)}^3}{R_{(i,j)}^1})$ | 0.01 | 10.07 | 0.6264 | 0.5996 |
| $a(\frac{S_i^2 \times V_{(i,j)}^3}{R_{(i,j)}^2})$ | 0.009 | 12.74 | 0.5271 | 0.6746 |
| $a(\frac{S_i^2 \times V_{(i,j)}^3}{R_{(i,j)}^3})$ | 0.009 | 14.38 | 0.4665 | 0.7166 |
| $a(\frac{S_i^3 \times V_{(i,j)}^1}{R_{(i,j)}^1})$ | 0.02 | 14.37 | 0.4669 | 0.7163 |
| $a(\frac{S_i^3 \times V_{(i,j)}^1}{R_{(i,j)}^2})$ | 0.02 | 15.31 | 0.432 | 0.7394 |
| $a(\frac{S_i^3 \times V_{(i,j)}^1}{R_{(i,j)}^3})$ | 0.02 | 16.22 | 0.3982 | 0.7611 |
| $a(\frac{S_i^3 \times V_{(i,j)}^2}{R_{(i,j)}^1})$ | 0.009 | 11.68 | 0.5664 | 0.646 |
| $a(\frac{S_i^3 \times V_{(i,j)}^2}{R_{(i,j)}^2})$ | 0.009 | 13.56 | 0.497 | 0.6958 |
| $a(\frac{S_i^3 \times V_{(i,j)}^2}{R_{(i,j)}^3})$ | 0.008 | 14.86 | 0.4485 | 0.7286 |
| $a(\frac{S_i^3 \times V_{(i,j)}^3}{R_{(i,j)}^1})$ | 0.003 | 10.78 | 0.5998 | 0.6206 |
| $a(\frac{S_i^3 \times V_{(i,j)}^3}{R_{(i,j)}^2})$ | 0.003 | 13.12 | 0.513 | 0.6846 |

Tables 5-7 shows the results when our model for calculated privacy risk $P_{(i,j)}$ was tested against the perceived privacy risk $F_d$. In these tables SSE : Sum of Squares due to Error, $R^2$ : Square of the correlation between the calculated $P_{i,j}$ and the observed $P_{i,j}$, and RMSE : Root Mean Squared Error. Table 5 shows that when we give the same power to all three parameters in the relationship the error is relatively high with a low R-square value. In table 6, we tried all 27 combinations of the powers 1,2 and 3 for S,V,R combinations without the combinations where all parameters have the same power. That is we ignored the combinations (1,1,1), (2,2,2) and (3,3,3). Table 6 shows that the goodness of fit increases with the increase of the power of visibility and decreases when the power of sensitivity and relatedness increases. Therefore, in table 8 we gradually increased the power of visibility and tested the goodness of fit while keeping the power of sensitivity and relatedness at 1.

**Table 7. Model Fitting - increasing visibility**

| Model | a (95% CI) | Goodness of fit | | |
|---|---|---|---|---|
| | | SSE | $R^2$ | RMSE |
| $a(\frac{S_i^1 \times V_{(i,j)}^4}{R_{(i,j)}^1})$ | 0.01 | 7.872 | 0.7079 | 0.5302 |
| $a(\frac{S_i^1 \times V_{(i,j)}^5}{R_{(i,j)}^1})$ | 0.003 | 7.723 | 0.7134 | 0.5252 |
| $a(\frac{S_i^1 \times V_{(i,j)}^6}{R_{(i,j)}^1})$ | 0.001 | 7.682 | 0.7149 | 0.5238 |
| $a(\frac{S_i^1 \times V_{(i,j)}^7}{R_{(i,j)}^1})$ | 0.01 | 7.682 | 0.715 | 0.5238 |
| $a(\frac{S_i^1 \times V_{(i,j)}^8}{R_{(i,j)}^1})$ | 0.01 | 7.693 | 0.7145 | 0.5242 |
| $a(\frac{S_i^1 \times V_{(i,j)}^9}{R_{(i,j)}^1})$ | 4.378e-05 | 7.706 | 0.7141 | 0.5246 |

We can see that the error increases again the power of visibility increases beyond 7. Therefore, the optimal relationship with the best goodness of fit is in the model where visibility is raised to the power of 7 with a coefficient of 0.01. This had a SSE of 7.6 and an $R^2$ of 71.5%. However the increase of $R^2$ from the model with visibility to the power three to visibility to the power 7 is only almost 1%. Therefore, one could safely assume that the model,

$$\frac{0.03 \times S_i \times V_{(i,j)}^3}{R_{(i,j)}}$$

gives a good enough approximation of the privacy risk perceived by users for a data item $i$ in a software application $j$. From the results, it is apparent that the visibility has the largest effect on the perceived privacy risk. In order to further observe why users felt differently when they disclosed data in the three scenarios we used in the study, we present the qualitative analysis of the reasons users gave.

### 5.1. Qualitative Analysis

Table 8 summarizes the codes we generated through the qualitative analysis. When it comes to the properties of data, participants mentioned that sensitivity, relevance and visibility of the data items affected their disclosure decisions. However, from their answers we could not identify any other attribute related to the data itself that affected the privacy risk perceived by users when they disclosed data. Participants mostly mentioned relevance of data (26%) followed by sensitivity of data (15%) and data visibility (12%). Nevertheless, our model showed that the visibility of data had the highest impact on the privacy risk perceived by users. Concerning visibility, participants said *If the application provides some tools to hide private information from public, it is fine* and *the controls on the data we disclosed are important*.

Consequently, we identified that users are concerned about the trust towards the organization that develop and publish applications (19%). Participants said that they are comfortable sharing data as long as the application is developed and owned by a trusted organization. This explains the relatively low mean perceived privacy risk we observed for the banking app, probably because users trusted their bank more. When it comes to trust, some participants spoke about the trust with the application itself rather than the organization (11%). Interestingly, participants said that they build trust based on common sense, *Thats due to the feeling of trust I have with them. i'm aware i should read information disclosure agreement. But I'm not reading it most of the time and use common sense*. This is an interesting finding that could be investigated further to see how users build trust with applications without reading the privacy policy. Some participants also raised concerns about personal safety (12%). Their concerns on personal safety was two fold. One, financial and reputation loss when data is accessed by unknown parties and two, being subjected to unwanted marketing via phone and email. They said they consider being exposed to unwanted advertising as a personal threat. A small number of participants mentioned their personal experience, the news they hear and also the benefits they could gain through data disclosure.

### 6. Discussion

The model we propose in this research is derived based on the theoretical knowledge presented by Maximilien et al. [8]. They propose that privacy risk could be measured by sensitivity (S) and visibility (V) where S and V are in any arbitrary relationship that results in a monotonically incremental result for privacy risk. However, their model has been applied on the assumption that both S and V of content has the same effect on the privacy risk [20]. Consequently, their model does not account for the relatedness of content. In our model we introduced a term for relatedness (R) of the content and through a user study we were able to identify that V had more impact on the privacy risk of the content than S and R. Our model shows that content visibility should be considered at a higher power to closely approximate the perceived privacy risk. This suggests that developers could significantly reduce the privacy risk perceived by users by controlling the visibility of their data within the system. That is in a system design, after measuring the privacy risk perceived by users against the data that is used in the system, developers could reduce the visibility of data with high privacy risk. When data is less visible in a system, the risk associated with those data reduces. This principle is also coined by unobservability and undetectability. Thereby suggesting that this would also reduce the actual privacy risk of data items in the system. Our model also shows that the R of data is in a monotonically decreasing relationship with users' perceived privacy. This suggest that developers should focus on using data that is absolutely necessary (higher relatedness) for the applications. Data privacy regulations such as the GDPR also emphasize this need [1]. Therefore, though reducing the perceived privacy risk of data, system developers can also reduce the actual privacy risk of data within their system designs.

For the categorization of data according to S, V and R we used three categories. In the workshop to determine S, V and R values with software developers, we encouraged the developers to further define categories if they felt three categories were not sufficient to handle the variations in S, V and R of data. We also asked them to challenge and argue on the definitions we have provided. While the participants agreed with three categories for V and R, they said that S may require more categories to identify sensitive data and extremely sensitive data. However, when they created one more category for extremely sensitive data, they ended up moving all data in the sensitive category to the extremely sensitive category and hence ending up with three categories at the end. Therefore, the participants agreed that the three categories we defined sufficiently captures the S, V R variation in data.

Our model provides developers with a measurable approach to understand users' perceived privacy risk. While previous research has always highlighted the need for software developers to understand and acknowledge user privacy requirements [3, 6], involvement of actual users in the system design process is not considered practical due to higher costs and time constrains [2]. Our model provides a cost effective alternative for developers to approximate the privacy risk perceived by users when they design software systems. Furthermore, compared to the soft measurements developers are expected to make in most scenarios that involve user privacy, we argue that this model to measure users' perceived privacy risk would be meaningful and pragmatic. For example, it has been previously coined that when implementing privacy in software systems, developers find it difficult to interpret the requirements to anonymize appropriate data, encrypt sensitive data, when they are required to make soft decisions which are not measurable [4]. The proposed model would help developers to understand data and the perceived privacy risk associated with data [26]. This knowledge could be used within existing privacy guidelines, to measure the privacy risk of data. Thereby identifying data considered as high risk and implement techniques in system designs to protect data. However, we do not go to the extent of arbitrarily proposing ways to use the model proposed here. Rather, we suggest that privacy engineers and system developers could utilize the knowledge presented in this paper to implement and protect user data in system designs, paving the way for designing privacy aware software systems.

Table 8. Issues participants faced when embedding privacy into the designs

| Code | Representative Quotes | Coverage |
| --- | --- | --- |
| Benefit to me | how it benefits myself/ how useful it is for me. | 2.64% (4) |
| How much I need the app | based on my requirements from the application | 7.2%(11) |
| News I see | by considering cyber crimes and all that | 0.66%(1) |
| Personal experience | I was in couple of these situations which gave me an idea | 2%(3) |
| Personal Safety | Some data could cause reputation and/or financial loss | 12% (19) |
| Relevance of data | if I don't think such applications needs the data | 26% (40) |
| Visibility of Data | whether I could control what others see | 12% (19) |
| Sensitivity of Data | some sensitive information can't be disclosed irrespective of the application | 15% (23) |
| Transparency | Depends on what they are going to do with the information | 6.6% (10) |
| Trusting the application | every online application cannot be trusted | 11% (17) |
| Trusting the organization | If it is a reputed or a government institution there is less doubt and more trust | 19% (29) |

Consequently, the model we derived here does not account for the human attributes of users that affect their perceived privacy risk when interacting with software systems. Previous research has shown that the personality of users affects the privacy risk they perceive when they interact with software systems. For example, Westin's privacy personality scale [27] shows that users could be divided into privacy fundamentalists (extremely concerned about privacy), privacy pragmatists (believe that privacy needs to be compromised according to situations) and privacy unconcerned, (little or no concern about privacy) [27]. Indicating the effect of such personalities, in our survey one participant said *Basically I feel comfortable giving information on a need to know basis only* and another one said *nothing* implying he did not feel different disclosing data into different application settings. This could be explained by the theory of psychometry, which explains why people's perception of external factors such as privacy is dependent on their psychological differences [11, 28]. There is a lot of work done in this area where privacy psychometry is scaled and defined. For example IUIPC is one such scale that defines how people differ in their privacy attitudes [11]. These scales suggest that attributes such as previous experience and the nature of work they do that may affect users' perceived privacy risk. For example, P5 said *With the experiences when surfing in the internet made me to answer above questions so* and P89 said *I was in couple of these situations which gave me an idea to answer these questions easily*. However, in this research our focus was to model the perceived privacy risk eliminating the personality traits of a person. Therefore, by design our survey did not capture the privacy profile of our participants. The model we tested had an SSE value of 7.682 and an $R_2$ value of 71%, which is an acceptable goodness of fit in a human study. While the variations in the model could probably be explained by human factors, for the purpose of deriving a model for software developers to approximate the privacy risk users perceive related to the data used in software systems, our model is appropriate. As future work, we aim to improve our study with privacy profiling of participants incorporating the models that capture psychometric measurements [11, 27, 28], in order to observe how our model could cater for users with different privacy personalities.

## 7. Conclusions

In this research we used the sensitivity of data, the visibility data gets in a system design and the relatedness of data to the system as the independent variables and proposed a model to measure users' perceived privacy risk based on existing theoretical knowledge. We then tested our model against the privacy risk perceived by users in three different application settings. Our results indicate that both sensitivity and visibility of content must be in a monotonically increasing combination to represent the perceived privacy risks. At the same time relatedness of the content should be in a combination with sensitivity and visibility such that privacy risk monotonically decrease with the relatedness. The model shows that content visibility has the highest impact on the perceived privacy risk of users.